\documentclass[epj]{svjour}
\usepackage{graphics}
\begin{document}
\title{Asymptotic dynamics of closed-boundary vehicular traffic}
\author{Chi-Lun Lee \and Chia-Ling Huang
}                     
%
%
\institute{Department of Physics, National Central University, Jhongli 32001, Taiwan
}
\date{Received: date / Revised version: date}
%
\abstract{
We study the dynamics of vehicular traffic in a loop using a car-following model with the consideration of volume exclusions. In particular, we solve the steady state for the single-cluster case and derive fundamental diagrams, exhibiting two branches representative of entering and leaving the jam, respectively. By simulations we also observe that the speed average over all vehicles reaches the same constant at the steady states, regardless of the final clustering state. The autocorrelation functions for the overall speed average and single vehicle speed are investigated, each revealing a robust time scale. The effects of noises in vehicular acceleration are discussed.
} 
\maketitle
\section{Introduction}
\label{intro}
The framework of statistical mechanics has reached acclaimed triumph over the last century. The remarkable triumph simply lies on its success for reducing complicated many-body problems of Hamiltonian dynamics into simple thermodynamic relations. This success leads to speculations of applying its framework towards other many-body problems, such as the granular system\cite{granularEdwards,granularHerrmann,granularLiu}, at which dissipation is introduced into the microscopically Newtonian dynamics.

Meanwhile, a great deal of attention has been paid on self-driven systems\cite{Viscek95,Schutz00,Schweitzer01}, with which it is much harder to build up the analogy to the framework of statistical mechanics. The traffic problem may provides an intriguing example\cite{Lighthill55,Montroll58,Prigogine71,NS92,Bando95,Intelligent00,Kerner04,Mahnke05,Sugiyama08,Kerner09}. In addition to the nonlinearity in its microscopic interactions, there are neither energy nor momentum conservations in a traffic system. Furthermore, the Newton's third law fails also, as each vehicle often makes responses according to the behavior of its neighboring vehicle in front instead of the one behind. These features makes the problem almost inaccessible from the routes of traditional statistical mechanics. Nevertheless, the studies and observations from traffic dynamics reveal some features commonly shared with statistical mechanics, such as phase transitions and fluid-dynamic features.

In this work we aim to study the asymptotic dynamics and its dynamical fluctuations in a closed-boundary region from our traffic model. It is worth noting that experiments have been performed with such a circular road setting by real vehicles\cite{Sugiyama08}. The asymptotic evolution can be compared to a thermodynamic equilibrium in the language of statistical mechanics, as the overall pattern remains steady while the `microscopic' objects keep on motioning if possible. For the traffic system, the asymptotic regime bears time-translational symmetry apart from minor fluctuations. Therefore macroscopic quantities such as the overall energy and momentum reach rather steady values at this regime.

The rest of this paper is organized as follows. In the next section we introduce the algorithms of our traffic model. The simulation results for the noise-free case are shown in Sec.~\ref{noise_free}, as it is shown that the system self-assembles into asymptotic steady states of various jamming patterns. Prompted by the observation of a cyclic, repetitive structure in vehicular profiles, we solve for the steady state directly from the algorithms in Sec.~\ref{steady}, and derive fundamental diagrams and the retarded time thereby. The effects of noises in vehicular acceleration are studied from the autocorrelation functions introduced in Sec.~\ref{noise}, and there turns out to be two robust timescales in the autocorrelation functions . Our results are summarized in Sec.~\ref{summary}.

\section{Model}
\label{model}
In our current work, we choose to use a continuous car-following model instead of a cellular automaton model\cite{NS92}, because the former gives a better description about dynamics. In the simplest car-following model, the proportionate control\cite{Montroll58}, the acceleration of a vehicle is dependent on its velocity difference to the leading vehicle. However, based on our daily driving experiences, this correlation decays as its headway (the distance to the leading vehicle) increases, and at a large distance the vehicle tends to accelerate towards some optimal velocity $v_0$. Therefore we adapt a modified car-following model that interpolates between these two scenarios. Moreover, we put in an optional noise term for the acceleration of each vehicle, representing internal fluctuations of vehicles.

As to our algorithm, the acceleration of the $i$th vehicle in free traffic is:
\begin{equation}
  \frac{dv_i}{dt} = \lambda (v^{\mathrm{next}}_i - v_i) + \eta \, ,
  \label{accel}
\end{equation}
where
\begin{equation}
  v^{\mathrm{next}}_i \equiv v_{i+1} + (v_0 - v_{i+1})(1-e^{-\Delta x_i /D_f}) \, ,
  \label{vnext}
\end{equation}
as $D_f$ is the characteristic car-following distance, $\Delta x_i$ is the distance to the leading vehicle, and $\eta$ is an optional noise term representing the internal fluctuations of each driver. In addition to the modified car-following algorithm as Eq.~\ref{accel}, we put on an extra braking algorithm to enforce the volume-exclusion criterion $\Delta x_i \geq D_c$, where $D_c$ is the length of each vehicle, as in this work we assume the vehicles to have a uniform size. Thus if at one moment the profile shows $\Delta x_i = D_c$ for some $i$ and the projected update from Eq.~\ref{accel} results in a violation of volume exclusion, then we reset $v_i=0$. This algorithm mimics our driving experience in the sense that a driver tends to stop to avoid an accident if the leading vehicle is in very close proximity. At last, we require that a vehicle at stop will start to accelerate according to the algorithm in Eq.~\ref{accel} only if its distance to the leading vehicle is larger than some safety distance $D_s$.

From the accelerating algorithm in Eq.~\ref{accel}, our model bears resemblance to the intelligent driver model\cite{Intelligent00}, which also takes into consideration the dependence on both the velocity difference and distance to the leading vehicle. Furthermore, the consideration of a noise term and the volume-exclusion criterion is characteristic in the Nagel-Schreckenberg model\cite{NS92}. Thus one can treat our model as a continuous version of the Nagel-Schreckenberg model with more emphasis on the acceleration and deceleration dynamics. Also the enforcement of the volume-exclusion criterion points out an extra decelerating mechanism which is much faster than the usual car-following mechanism, although in our model this extra deceleration is performed instantly.

In this work we apply a periodic boundary condition, so that the vehicles can be treated as proceeding on a ring of length $L$. Unless mentioned we set the vehicle length $D_c$=3 meters, safety distance $D_s$=6 meters, the characteristic following distance $D_f$=60 meters, and $L$=1000 meters. We set the optimal velocity $v_0 = 25$m/s = 90km/hr. The parameter $\lambda$ represents the pace of car-following adaptation, and the larger $\lambda$ is, the sooner the vehicle $i$ adjusts to $v^{\mathrm{next}}_i$. In this work we set $\lambda=0.15$ s$^{-1}$, such that it takes about 20 seconds for a vehicle at stop to reach velocity $0.95 v_0$ if its front is empty of other vehicles. During our simulations we set the time interval between successive updates to be $\Delta t=0.001$s. We use the SI units throughout this report. As to the initial conditions for all of our simulations, we set the vehicles to distribute uniformly over the road, while we use various speed profiles.

\section{Simulation results: noise-free case}
\label{noise_free}
\begin{figure}
  \includegraphics{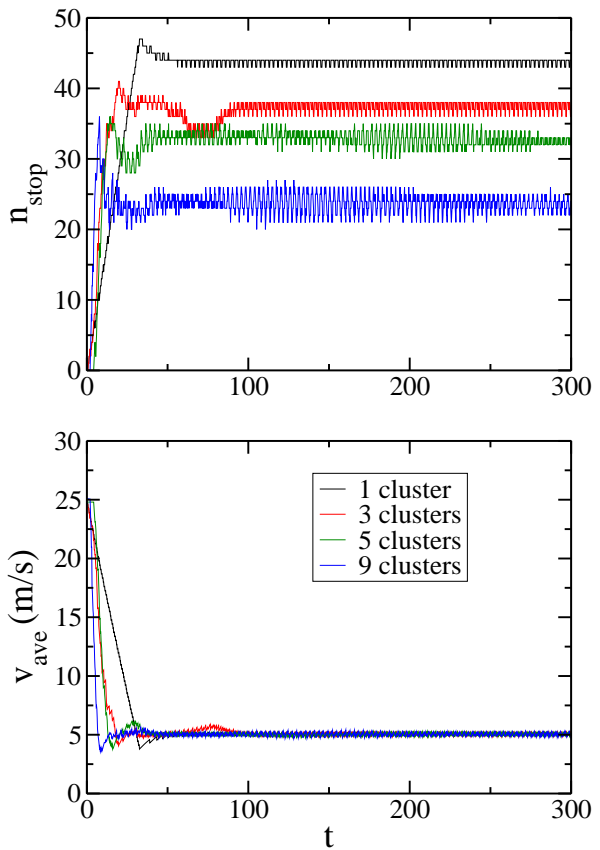}
  \caption{Time evolution for $n_\mathrm{stop}$ , the total number of stopped vehicles, and $v_\mathrm{ave}$, the speed average over all vehicles. The curves show results of trials over different initial speed distribution with the setting $D_s=6$m. The curves are labeled by the number of jammed clusters observed at their asymptotic steady states.}
  \label{n_v_safe}
\end{figure}

\begin{figure}
  \includegraphics{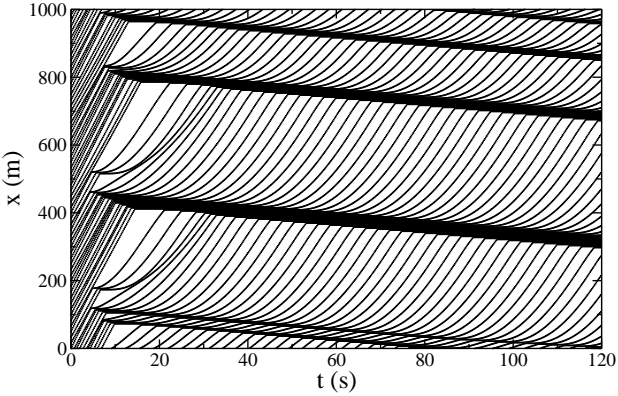}
  \caption{The trajectories of vehicles during one simulation. At large time the vehicles tend to repeat a single pattern cyclically, as each vehicle repeats the behavior of its leading vehicle with a time delay.}
  \label{pos_no_noise}
\end{figure}

Fig.~\ref{n_v_safe} shows the time evolution in our simulations for various initial conditions, as $n_\mathrm{stop}$ is the number of vehicles stopped and $v_\mathrm{ave}$ is the speed average over all vehicles. Although it is well known that the algorithm of proportionate control does lead to stable traffic asymptotically\cite{Montroll58}, the algorithm itself does not guarantee that vehicles do not overtake their front neighbors. With the extra consideration of volume exclusion, jams occur due to the propagation of the sudden-brake behavior. For our closed-boundary system, the propagation of jamming often comes to a stop upon reaching a depleted zone, as the traffic dynamics finally evolves towards an asymptotic steady state with some rather stabilized pattern. At the asymptotic steady state the quantities $n_\mathrm{stop}$ and $v_\mathrm{ave}$ exhibit slight, regular fluctuations only. In our simulations we have not observed any chaotic speed profiles asymptotically for the noise-free case. Fig.~\ref{pos_no_noise} shows an example in which the asymptotic steady state consists of five jammed clusters. One observes that at large time the vehicles tend to repeat a single pattern cyclically, as each vehicle repeats the behavior of its leading vehicle with a fixed time delay.

It is worth noting that Fig.~\ref{n_v_safe} shows the speed average $v_\mathrm{ave}$ reaches approximately the same asymptotic value despite the various initial conditions considered. Therefore no matter how many jammed clusters it forms, the speed average of the system stays the same asymptotically, apart from minor fluctuations. The only exception happens for the non-jammed state, in which $v_\mathrm{ave} = v_0$. This is less intuitive as we often think that $v_\mathrm{ave}$ would be smaller if more jams are encountered at the traffic. This fact can be helped understood partially noting that in Fig.~\ref{n_v_safe}, one finds less total number of stopped vehicles instead as there are more jammed clusters. However, it remains to account for the remarkable coincidence that $v_\mathrm{ave}$ reaches the asymptotic same value regardless of jamming patterns.

\begin{figure}
  \includegraphics{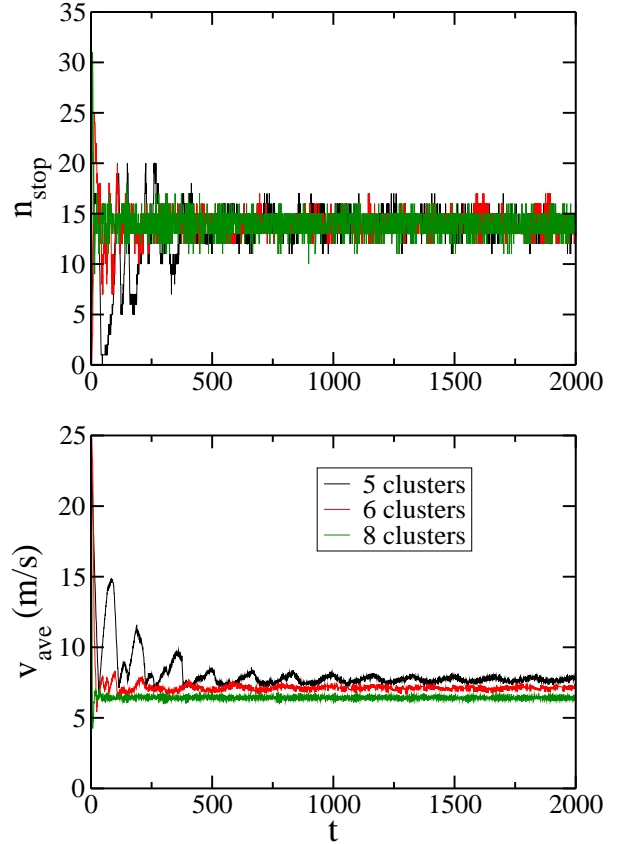}
  \caption{Time evolution for $n_\mathrm{stop}$ and $v_\mathrm{ave}$ for the case $D_s=0$m. The curves show results of trials over different initial speed distribution, as curves are labeled by the number of jammed clusters observed at their asymptotic steady states.}
  \label{n_v_nosafe}
\end{figure}

As a comparison, we set the safety distance $D_s$ to zero and find that $v_\mathrm{ave}$ does not reach a fixed value over different jamming pattern formations. Fig.~\ref{n_v_nosafe} shows the time evolution of $n_\mathrm{stop}$ and $v_\mathrm{ave}$ for two examples. Furthermore, we also find that it takes a much longer time for the fluctuations to die out before the system reaches an asymptotic steady state. Also the lack of a safety distance implies that large fluctuations could arise at a slight perturbation. For example, if we set that initially $v_i(t=0) =v_0$ for all $i$ except $v_1 = v_0-20$m/s. While for the case with $D_s=6$m we observe a single cluster pattern, a four-cluster pattern emerges for the case $D_s=0$m, which suggests that without a safety distance, subsequent jams can be triggered easily from a local disturbance.

\section{Steady-state solution of single-cluster formation}
\label{steady}
From our simulation results for the noise-free case, we find at as time proceeds, the traffic evolves towards a steady pattern, as is evidenced by the stabilized macroscopic quantities, such as the averaged speed and number of stopped vehicles. Furthermore, the position profile of vehicles reaches a cyclic pattern asymptotically. Assuming that asymptotically all the fluctuations and irregularities turn to die out, and vehicles are proceeding with same velocity pattern except with time shifts, one can solve for the asymptotic behavior from our car-following algorithms described in Sec.~\ref{model}.


Let us define $t=t_{\mathrm{min}}$ to be the time that the $i$th vehicle starts to dissolve from the jam, and $t=0$ to be the time for it to reenter the jam. Therefore one has
\begin{equation}
  \Delta x(t=0) = x_{i+1}(0) - x_i(0) = D_c \, ,
  \label{BC_Dc}
\end{equation}
and
\begin{equation}
  \Delta x(t=t_{\mathrm{min}}) = D_s \, .
  \label{BC_Ds}
\end{equation}
From the cyclic pattern we assume that at the steady state the velocity profile of each vehicle shifted from that of its leading vehicle simply by a time delay $\tau$, i.e.,
\begin{equation}
  v_i(t) = v_{i+1}(t-\tau) \, .
  \label{recurrent}
\end{equation}
By putting Eqs.~\ref{vnext} and \ref{recurrent} into Eq.~\ref{accel}, we get
\begin{eqnarray}
  \frac{1}{\lambda} \frac{dv_i}{dt} &=& v_{i+1}(t) - v_i(t) + [v_0 - v_{i+1}(t)] \cdot [1-e^{-\Delta x_i(t) /D_f}] \nonumber \\
  &=& v_0 - v_i(t) - [v_0 - v_i(t+\tau)] \nonumber \\
  & & \ \ \cdot \left\{\exp \left[ -\frac{D_c + \int_0^t \! dt' \, [v_i(t'+\tau)-v_i(t')]}{D_f} \right] \right\}  ,
  \label{eqn_steady}
\end{eqnarray}
where we have used the fact that
\begin{eqnarray}
  \Delta x_i(t) &=& \int_0^t \! dt' \, [v_{i+1}(t') - v_i(t')] + \Delta x_i(0) \nonumber \\
  &=& \int_0^t \! dt' \, [v_i(t'+\tau) - v_i(t')] +D_c \nonumber \\
  &=& \int_{t}^{t+\tau} \! dt' \, v_i(t') - \int_0^{\tau} \!dt' \, v_i(t') +D_c \nonumber \\
  &=& \int_{t}^{t+\tau} \! dt' \, v_i(t') +D_c \, .
  \label{deltax}
\end{eqnarray}
Note that we assume that $v_i(t) = 0$ for $t>0$. This assumption holds as long as each vehicle comes to a stop before its leading one starts. Eq.~\ref{eqn_steady} is an integro-differential equation that cannot be solved by simple numerical integration. Since the delayed time $\tau$ is itself yet to be known, we resort to an iteration procedure instead. To determine the value of $\tau$, we need an extra identity, which is given by Eq.~\ref{BC_Ds} (along with the use of Eq.~\ref{deltax}):
\begin{equation}
  \Delta x_i(t_{\mathrm{min}}) = D_s = \int_{t_\mathrm{min}}^{t_{\mathrm{min}}+\tau} \! dt' \, v_i(t') + D_c \, .
  \label{constraint_tau}
\end{equation}

Note that the boundary condition for Eq.~\ref{eqn_steady} is $v_i(t=t_{\mathrm{min}})=0$. The value of $v_i$ before reentering the jam $v_i(0^{-})$ will be determined upon solving Eq.~\ref{eqn_steady}. Thus each iteration consists of two steps. First we decide a new value of $\tau$ to match the identity in Eq.~\ref{constraint_tau} using the previous speed profile. Then we derive the updated speed profile according to Eq.~\ref{eqn_steady}. In practice we find it very effective to rewrite Eq.~\ref{eqn_steady} as
\begin{equation}
  e^{-\lambda t} \frac{d}{dt}[e^{\lambda t}(v_i(t) -v_0)] = \lambda [v_i(t+\tau) - v_0] e^{-\Delta x_i(t)/D_f} \, ,
  \label{eqn_steady2}
\end{equation}
and for each iteration we plug in the old speed profile on the right-hand side and derive the updated profile from the left-hand side of Eq.~\ref{eqn_steady2}. By this algorithm we can get a steady-state solution with the sums of residual squares less than $\times 10^{-4}$ within 15 iterations, as we set the time interval between successive updates to be $\Delta t=0.001$. Before iteration starts we set our initial guess of the speed profile to be $v_i(t) = v_0 \cdot \{1-\exp[-\lambda (t-t_{\mathrm{min}})]\}\cdot [1-\exp(\lambda t)]$.

\begin{figure}
  \includegraphics{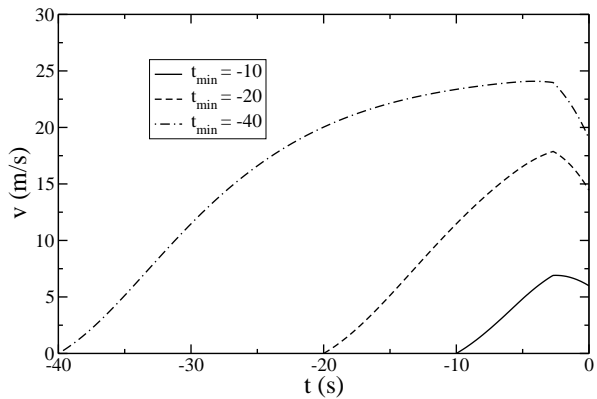}
  \caption{Speed curves of steady state traffic for $t_\mathrm{min}=-40$, -20, and -10 seconds.}
  \label{v_t}
\end{figure}

In Fig.~\ref{v_t} the steady-state solutions of the speed profiles for the cases $t_\mathrm{min}=-40$, -20, and -10 seconds. We stress again here that they are derived from the single-cluster steady-state condition. Note that $-t_\mathrm{min}$ indicates the time of free traffic for a single vehicle between successive stops. One can observe a sharp, discontinuous change in the acceleration of the vehicle, which occurs at the moment when its leading vehicle stops. Furthermore, there exist great asymmetry between the accelerating and the decelerating parts of the speed profile, a result of the feature that there is no time-reversal symmetry in the microscopic traffic dynamics. The second discontinuity in Fig.~\ref{v_t} happens at $t=0$, when the speed drops to zero abruptly. This implies that the car-following algorithm we used in Eq.~\ref{accel} cannot guarantee a collision-free environment and the driver has to resort to an extra, instantaneous braking behavior to maintain the volume exclusion principle.

\begin{figure}
  \includegraphics{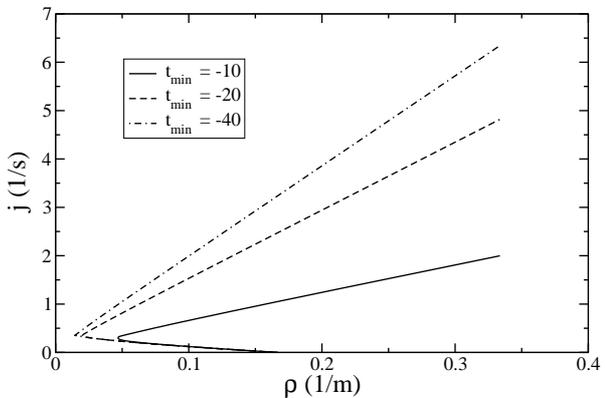}
  \caption{Fundamental diagrams of steady state traffic for $t_\mathrm{min}=-40$, -20, and -10 seconds.}
  \label{jrho}
\end{figure}

From the steady-state solution we can obtain fundamental traffic diagrams as follows. First of all one has $j = \rho v$, where $j$ is the traffic flow and $\rho$ is the density of vehicles. Since we already have the speed profile $v_i(t)$, we can use the definition that $\rho(t)=1/\Delta x_i(t)$, the inverse of the spacing between successive vehicles. Therefore we get $j(t) = v_i(t)/\Delta x_i(t)$, as the fundamental diagram is plotted in Fig.~\ref{jrho} by collecting the serial data of $(\rho(t), j(t))$. For each $t_\mathrm{min}$ the fundamental diagram shows two branches. The upper branch represents the dynamics towards a jam, while the lower one shows the dynamics coming out of the jam. The results shows that the slope for the upper branch in the fundamental diagram changes significantly at small $|t_\mathrm{min}|$ as $t_\mathrm{min}$ modifies, while at larger $|t_\mathrm{min}|$ the curves merge toward a single one. Note that since the lower branch represents dynamics coming out of a jam, where the vehicles are accelerating greatly from stop, the curves at this branch are more sensitive to the definitions of $j$ and $\rho$.

One can estimate the receding velocity of the jamming front $v_\mathrm{jam}$ from the steady-state solution via two routes. First, realizing that for each time interval $\tau$, a vehicle enters into the jam while another leaves the jam, one concludes that $v_\mathrm{jam} = - D_c/\tau$. The second method follows from the derivation of the shock-wave velocity in Ref.~\cite{Lighthill55}. Hence $v_\mathrm{jam}$ can be estimated from Fig.~\ref{jrho} by the slope of the line connecting the points of the jammed state (where $j=0$) and the free state (estimated by the point with lowest density allowed in the fundamental diagram). For the case $t_\mathrm{min}=-40$ the both methods give $v_\mathrm{jam} \approx -1.11$ m/s. Furthermore, the value $v_\mathrm{jam}$ changes less significantly over $t_\mathrm{min}$. For example, our solution shows $v_\mathrm{jam} \approx -1.10$ m/s when $t_\mathrm{min}=-10$s.

The length of the free-traffic section at the steady state $L_\mathrm{free}$ can be evaluated in the following manner. Since $-t_\mathrm{min}$ is the duration of time for each vehicle between successive stops, one has
\begin{equation}
  L_\mathrm{free} = \int_{t_\mathrm{min}}^{0} \!\! dt \, v_i(t) - v_\mathrm{jam} \cdot \tau \, ,
  \label{L_free1}
\end{equation}
due to the fact that the jamming front moves backwards with velocity $v_\mathrm{jam}$ while the vehicle proceeds.
Alternatively, one can derive $L_\mathrm{free}$ by summing over the spacings between successive vehicles at $t=t_\mathrm{min}$:
\begin{equation}
  L_\mathrm{free} = \sum_{j=1}^{n} \Delta x_{i+j}(t_\mathrm{min}) = \sum_{j=1}^{n} \Delta x_i(t_\mathrm{min}+j\tau) \, ,
  \label{L_free2}
\end{equation}
where $n$ is the number of vehicles at free traffic:
\begin{equation}
  n_\mathrm{free} = \left[ -\frac{t_\mathrm{min}}{\tau} \right] \, .
\end{equation}

\begin{figure}
  \includegraphics{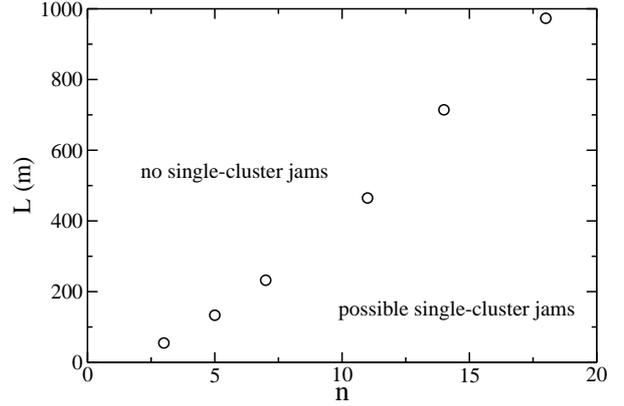}
  \caption{Phase diagram for single-cluster formation. $L$ is the circumference of the loop, and $n$ is the total number of vehicles.}
  \label{phase_diagram}
\end{figure}

Thus for each $t_\mathrm{min}$ one gets a corresponding $L_\mathrm{free}$ and $n_\mathrm{free}$ at the single-cluster steady state. Through this one-to one correspondence we can treat $L_\mathrm{free}$ as a function of $n_\mathrm{free}$. Therefore if $n$ vehicles are running on a loop of circumference $L > L_\mathrm{free}(n)$, then one cannot observe any single-cluster formation at the steady state. To verify this argument, we first note that the free-traffic section in single-cluster formations bears the property that the residual (empty) space at free traffic is a monotonically increasing of $n$:
\begin{equation}
  L_\mathrm{free}(n) - n D_c > L_\mathrm{free}(n') - n' D_c \ \ \ \ \mbox{if $n>n'$} \, .
  \label{mono_res_free}
\end{equation}
Now assume $n$ vehicles are distributed at a loop with circumference $L > L_\mathrm{free}(n)$. If at the steady state it forms a single cluster with $n-n'$ jammed vehicles, then the length of the free-traffic section is $L-(n-n')D_c$. Therefore we have
\begin{equation}
  L_\mathrm{free}(n') = L-(n-n')D_c \, .
  \label{assumption_free}
\end{equation}
However, Eq.~\ref{assumption_free} implies that
\begin{equation}
  L_\mathrm{free}(n') -n' D_c = L - n D_c > L_\mathrm{free}(n) - n D_c
  \label{contradiction}
\end{equation}
for some $n'<n$, which violates Eq.~\ref{mono_res_free}. Hence we conclude that for $n$ vehicles distributed at $L>L_\mathrm{free}(n)$ single-cluster formation does not exist at the steady state. Fig.~\ref{phase_diagram} shows such a phase diagram while $L_\mathrm{free}$ is evaluated using the second method. The results from the first method follow very closely, with residuals less than the size of a vehicle. Note that the condition that $L < L_\mathrm{free}(n)$ does not necessarily imply the existence of a jammed steady state, as states with a homogeneous speed distribution $v_i = v_0$ also act as stable solutions.

\section{The noise-added case: autocorrelation functions}
\label{noise}
Instead of putting random noises on all drivers at all times, we note that noises occur rather occasionally and choose put the noise term in our model as follows. During each time step $\Delta t$, each vehicle has a probability $p$ to suffer from random noise $\eta$ in its acceleration due to disturbance or else. The random variable $\eta$ bears a uniform distribution ranging between $-\eta_0$ and $\eta_0$. Throughout the course of simulations we use the value $\eta_0 = 1000 \mbox{m}/\mbox{s}^2$. Although the value of $\eta$ looks rather unphysical, within the time between successive updates $\Delta t=0.001$s we find that each noisy acceleration results in a change of a speed up to $\eta_0 \Delta t = 1$m/s, or equivalently, $3.6$km/hr.

For the noise-added case, the vehicles do not evolve into a cyclic, steady-state pattern asymptotically due to persistent internal fluctuations. Nevertheless, asymptotically the system still evolves into a stage resembling a thermal equilibrium. Hereby we study the nature of the asymptotic kinetics via autocorrelation functions. We define the autocorrelation function for the speed average $C_\mathrm{ave}(t)$ as
\begin{equation}
  C_\mathrm{ave}(t) \equiv \frac{\langle(v_\mathrm{ave}(t')-\langle v_\mathrm{ave} \rangle)(v_\mathrm{ave}(t'+t)-\langle v_\mathrm{ave} \rangle)\rangle}{\langle v_\mathrm{ave}^2 - \langle v_\mathrm{ave} \rangle^2 \rangle}\, ,
\end{equation}
and the autocorrelation for the speed of a certain vehicle $C_1(t)$ as
\begin{equation}
  C_1(t) \equiv \frac{\langle(v_i(t')-\langle v_i \rangle)(v_i(t'+t)-\langle v_i \rangle)\rangle}{\langle v_i^2 - \langle v_i \rangle^2 \rangle}\, ,
\end{equation}
as the bracket represents time average.

\begin{figure}
  \includegraphics{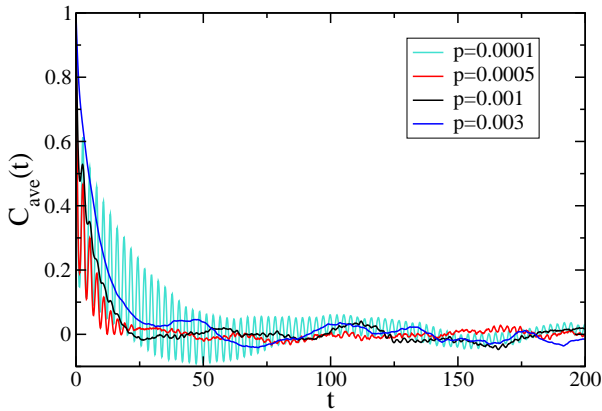}
  \caption{Autocorrelation function $C_\mathrm{ave}(t)$ of speed average $v_\mathrm{ave}$. The parameter $p$ represents the likelihood of noise occurrence.}
  \label{ACF}
\end{figure}

Fig.~\ref{ACF} shows the autocorrelation function $C_\mathrm{ave}(t)$ over various values of noise probability $p$. We find that for small $p$ the relaxation shows short-time oscillatory behavior, with the oscillation period close to $2.7$s. This oscillation period is approximately equal to the delayed time $\tau$, which we have derived from our steady-state analysis for single-cluster formation. Hence the short-time oscillations represent the synchronized behavior of vehicles entering and leaving jamming sections. We find that for the noise-free case the same oscillation period persists as well for steady states of multiple clusters. Furthermore, when the probability of noise occurrence $p$ increases, the oscillatory period also increases slightly. As $p$ further increases, the synchronized behavior eventually diminishes, and the short-time oscillations become absent in the relaxation of $v_\mathrm{ave}$, and the long-term memory possessed by the system in the noise-free case eventually gets destroyed, as $C_\mathrm{ave}(t)$ reaches a short-time exponential decay at large $p$.

\begin{figure}
  \includegraphics{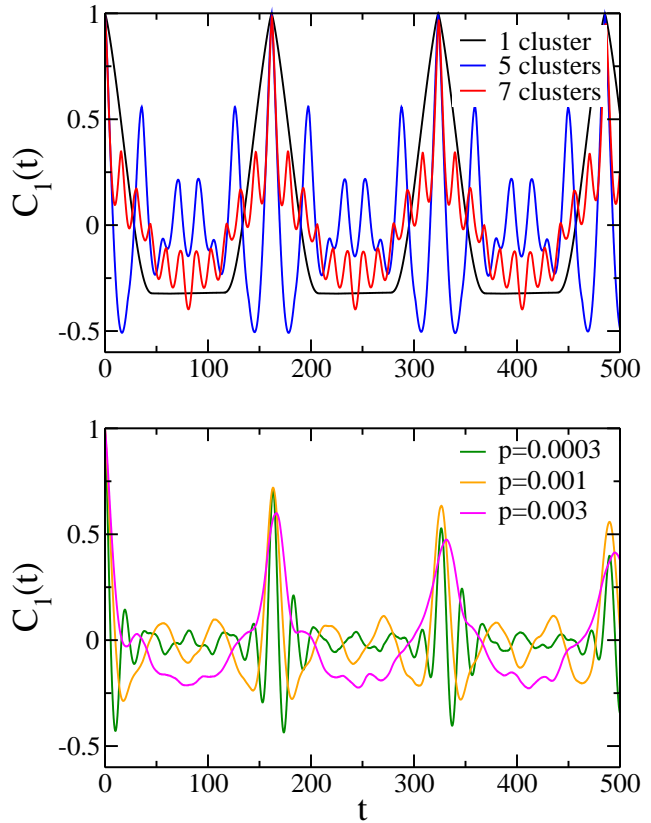}
  \caption{Autocorrelation function $C_1(t)$ of single-vehicle speed $v_i$. The parameter $p$ represents the likelihood of noise occurrence. Upper: noise-free cases ($p=0$); Lower: with added noises. The noise-free curves are labeled by the number of clusters formed at the steady states.}
  \label{ACFone}
\end{figure}

The relaxation of single-vehicle speed fluctuation $C_1(t)$ is presented in Fig.~\ref{ACFone}. For the noise-free case we observe peaks of height one for every time interval of approximately $162$s. By checking our single-cluster simulation result we find this is the time required for a vehicle to repeat its speed profile at the steady state. Furthermore, by examining our results over several noise-free cases we find the same time interval persists even for systems with multiple jammed clusters. Meanwhile, for multi-cluster cases the autocorrelation function $C_1(t)$ exhibits smaller peaks in between. When noises are added, we find this correlation to be rather robust, compared with the collective relaxation of $v_\mathrm{ave}$. The peaks remain significant even at $p=0.01$, while a more detailed check suggests the time interval between successive peaks slightly widens as $p$ increases.

For the noise-free case, the speed average $v_\mathrm{ave}$ exhibits small, regular fluctuations due to the repetitive stopping and accelerating motions of vehicles at jamming boundaries. When the noise level increases this regular fluctuation will be easily overpowered because of its small amplitude. Meanwhile, the structure or the pattern of the traffic are less susceptible to the noises. It thus explains why the short-time oscillations in $C_\mathrm{ave}(t)$ diminish quickly with noise increasing, while the major peaks in $C_1(t)$ remain up to a relatively high noise level. However, it remains to understand why the time intervals between peaks stay unchanged or are only slightly modified over different jamming patterns and noise levels.

One may note from Fig.~\ref{ACFone} that for the noise-free case the autocorrelation functions $C_1(t)$ exhibit symmetric patterns between successive largest peaks. This symmetry results from the fact that at the asymptotic state time-reversal symmetry holds for the autocorrelation function: $C_1(t) = C_1(-t)$, due to the fact that time-translational symmetry holds asymptotically: $\langle v(t') v(t'+t)\rangle = \langle v(t'-t) v(t')\rangle$. And for the noise-free case the system reaches a steady state asymptotically with a cyclic pattern. Therefore $C_1(t) = C_1(t+T) = C_1(-t) = C_1(T-t)$, where $T$ is the cyclic period. On the other hand, as the noises turn on, the autocorrelation function $C_1(t)$ fails to be strictly periodic, as asymmetric pattern emerges therein.

\section{Summary}
\label{summary}
It is important to understand the many-body dynamics of self-driven objects, which appears so often in biological and even social systems. The highway traffic dynamics provides a simple platform towards understanding the possible common behavior of self-driven systems. Although the microscopic dynamic equations are highly nonlinear compared with Newtonian dynamics, the traffic system with closed boundaries evolves towards a steady state for the noise-free case. We observe from simulations that the speed average over all vehicles eventually reaches approximately the same value, regardless of the clustering patterns. This remarkable feature is absent if we do not allow any safety distance before stopped vehicles start moving.

The steady states can be solved knowing that successive vehicles are following same speed profile with some retarded time $\tau$. And therefore one can obtain fundamental diagrams for the noise-free case. Moreover, we have derived a criterion for possible single-cluster jam formation.

Meanwhile, as noises are introduced for individual vehicles, the asymptotic evolution of the system can be analyzed via autocorrelation functions. The autocorrelation of the speed average $C_\mathrm{ave}(t)$ exhibits rapid oscillations with a period characteristic of the retarded time $\tau$. Again this period is robust over different traffic patterns, and this short-time regular oscillations for the speed average will be overpowered as the likelihood for noise occurrence $p$ increases. Furthermore, the single-vehicle speed autocorrelation function $C_1(t)$ shows periodic behavior with slower oscillations for noise-free cases. The period observed corresponds to the time for a single vehicle to repeat its speed profile. Again this timescale is robust over our various simulation trials. When noises are added we find these periodic peaks can live up to a high noise level.

%
%
%

\end{document}